\documentclass[12pt]{article}
\pdfoutput=1 
\usepackage{amsmath,amssymb,amsfonts,amsthm}
\usepackage{graphicx}
\usepackage{xcolor, colortbl}
\usepackage{empheq}
\usepackage[pdftex, bookmarks=true,colorlinks,linkcolor=red,urlcolor=blue, citecolor=blue]{hyperref}
\usepackage{cite}
\usepackage{subcaption}
\usepackage{slashed}

\usepackage{tikz}
\usepackage{amsthm}
\usepackage{array}
\usepackage{soul}
\usepackage{float}
\usepackage{bm}

\makeatletter\@addtoreset{equation}{section}\makeatother

\newcommand{\preprint}[1]{\begin{table}[t]  
             \begin{flushright}               
             {#1}                             
             \end{flushright}                 
             \end{table}}                     
\renewcommand{\title}[1]{\vbox{\center\LARGE{#1}}\vspace{5mm}}
\renewcommand{\author}[1]{\vbox{\center#1}\vspace{5mm}}
\newcommand{\address}[1]{\vbox{\center\em#1}}

\newcommand{\be}{\begin{equation}}
\newcommand{\ee}{\end{equation}}
\newcommand{\bea}{\begin{eqnarray}}
\newcommand{\eea}{\end{eqnarray}}
\newcommand{\bse}{\begin{subequations}}
\newcommand{\ese}{\end{subequations}}
\newcommand{\beqa}{\begin{eqnarray}}
\newcommand{\eeqa}{\end{eqnarray}}
\newcommand{\beqar}{\begin{eqnarray*}}
\newcommand{\eeqar}{\end{eqnarray*}}
\newcommand{\bi}{\begin{itemize}}
\newcommand{\ei}{\end{itemize}}
\newcommand{\bn}{\begin{enumerate}}
\newcommand{\en}{\end{enumerate}}

\newcommand{\ba}{\begin{array}}
\newcommand{\ea}{\end{array}}
\newcommand{\bc}{\begin{center}}
\newcommand{\ec}{\end{center}}

\definecolor{darkgreen}{rgb}{0,0.3,0}
\definecolor{darkblue}{rgb}{0,0,0.3}
\definecolor{darkred}{rgb}{0.7,0,0}

\def\lae{\mathrel{\mathop{\smash{\lower .5 ex \hbox{$\stackrel<\sim$}}}}}
\def\lae{\mathrel{\mathop{\smash{\lower .5 ex \hbox{$\stackrel>\sim$}}}}}

\def\arXiv#1{\href{http://arxiv.org/abs/#1}{arXiv:#1}}
\def\arXiv#1#2{\href{http://arxiv.org/abs/#1}{arXiv:#1}}

\def\doi#1#2{\href{http://doi.org/#1}{#2}}

\newcommand{\comment}[1]{{\bf {\textcolor{blue}{ [#1]}}}}

\begin{document}

\unitlength = .8mm

\begin{titlepage}
\vspace{.5cm}
\preprint{}

\date{\today}
\begin{center}
\hfill \\
\hfill \\
\vskip 1cm

\title{\boldmath Shear subdiffusion in non-relativistic holography 
}
\vskip 0.5cm
{Yan Liu$^a$}\footnote{Email: {\tt yanliu@buaa.edu.cn}}, 
{Zhi-Ling Wang$^a$} 
{and Xin-Meng Wu$^{b,c}$}\footnote{Email: {\tt xinmeng.wu@sjtu.edu.cn}}

\address{$^{a}$
Department of Space Science and Peng Huanwu Collaborative Center\\ for Research and Education, Beihang University, Beijing 100191, China}

\address{$^{b}$Wilczek Quantum Center, School of Physics and Astronomy, Shanghai Jiao Tong University, Shanghai 200240, China}

\address{${}^{c}$Shanghai Research Center for Quantum Sciences, Shanghai 201315, China}

\end{center}
\vskip 1.5cm


\abstract{
We study shear fluctuations in  non-relativistic holographic systems coupled to torsional Newton-Cartan geometry, using asymptotically Lifshitz spacetimes in Einstein-Maxwell-dilaton gravity. We identify a universal subdiffusive shear mode characterized by the quartic dispersion relation $\omega=-iD_4 k^4$, in sharp contrast to the conventional hydrodynamic diffusion. We derive this result analytically through a  systematic higher-order matched asymptotic expansion connecting near-horizon and far-region solutions, and we verify it with direct numerical quasinormal mode calculations. Our numerics demonstrate that the first non-hydrodynamic mode is purely imaginary and gapped, following the dispersion relation $\omega=-i\omega_0-i D k^2$, and that both the hydrodynamic and the first non-hydrodynamic modes pass through pole-skipping points. These results highlight  Lifshitz holography as an efficient framework for anomalous  transport in strongly coupled non-relativistic quantum matter.   
}

\end{titlepage}

\begingroup 
\hypersetup{linkcolor=black}
\tableofcontents
\endgroup

\vspace{0.5cm}
\section{Introduction}

Diffusion is a fundamental physical process that transports particles, energy, or other quantities from regions of high concentration to low concentration, driving systems toward equilibrium  \cite{diffusion-review}. 
Typically, it is described by Fick's laws $\omega=-iD k^2$, which follows  from conservation laws combined with  a diffusive flux proportional to  the concentration gradient. 
However, not all conserved quantities obey standard 
Fickian diffusion at late times.  
When microscopic dynamics involves 
additional symmetries, long-lived  memory effects or ergodicity breaking, 
the macroscopic Fickian description 
breaks down. In such cases, relaxation is governed by  subdiffusive hydrodynamic poles, leading to different scaling laws for the relaxation time.  

Subdiffusive transport 
plays a crucial role in 
understanding anomalous dynamical behaviors of many body systems. 
A canonical example arises in fluids with simultaneous conservation of charge and dipole moment, 
where enhanced symmetry constraints suppress microscopic kinetics, leading to subdiffusive charge transport \cite{Feldmeier:2020xxb,Gromov:2020yoc,Glodkowski:2022xje}. Further hydrodynamic realizations, such as those in tilted lattice systems, further illustrate the generality of subdiffusive phenomena  \cite{Zhang:2020der}. 
These unconventional subdiffusive dispersion relations signal exotic late-time dynamics in non-equilibrium systems, where relaxation rates follow unconventional scaling laws \cite{Guardado-Sanchez:2019bjm}. 
While most studies have focused on the subdiffusion of conserved charges,  which typically exhibits longitudinal subdiffusive behaviors, ---  momentum, 
a conserved quantity that normally follows shear (viscous) diffusion in the transverse direction, can also exhibit slower subdiffusive scaling under sufficiently strong constraints. 
 For example, in the presence of the magnetic field, the Lorentz force can  balance the longitudinal pressure gradient against the transverse friction force, causing the net force density to vanish at leading order and thereby inducing subdiffusive 
 momentum transport \cite{Hartnoll:2007ih}. Compared to charge or energy transport, momentum dynamics is often more challenging to study directly, making holographic methods particularly valuable.

In this work, we analyze non-relativistic strongly coupled field theories coupled to torsional Newton-Cartan (TNC) geometry using holography. Such field theories 
widely exist in condensed matter physics and we focus on a class of strongly coupled field theories dual to  Lifshitz black holes \cite{Taylor:2015glc}.  

Holographic duality is a powerful tool for studying strongly coupled systems by mapping them to a tractable weakly coupled gravitational system in a higher-dimensional curved spacetime \cite{Zaanen:2015oix,Hartnoll:2016apf}.
In the asymptotically AdS case, i.e. the celebrated AdS/CFT correspondence, 
this duality naturally yields standard Fickian diffusion in the hydrodynamic limit 
\cite{Policastro:2002se, Kovtun:2005ev}. 
This universality highlights the robustness of hydrodynamics but also reflects potential  limitations arising from the relativistic UV structure. Realizing exceptional subdiffusion in holography is therefore both challenging and interesting.  Subdiffusion has previously been achieved in certain holographic settings, including holographic magnetic hydrodynamics \cite{Hartnoll:2007ih, Ahn:2022azl},  a phenomenological holographic bulk action including higher form fields \cite{Ganesan:2020wvm} and  holographic solids  \cite{Xia:2025ewv}. However, the complexity of these constructions motivates us to explore  subdiffusion in a simpler non-relativistic holographic framework which is beyond the conventional  AdS/CFT:  Lifshitz holography. 

Lifshitz holography aims to describe the non-relativistic field theories, by introducing a Lifshitz scaling symmetry at the UV boundary \cite{Kachru:2008yh}. 
On the gravity side, Lifshitz spacetime can be constructed in bottom-up approaches, notably via Einstein-Maxwell-dilaton (EMD) theory with a  running scalar \cite{Taylor:2008tg}, or  Einstein-Proca theory with a massive vector field in the bulk \cite{Kachru:2008yh}. 
Due to the breaking of Lorentz boost, the non-relativistic Newton-Cartan structure in the dual field theory is involved. It turns out that the dual field theory Lifshitz black hole is a non-relativistic field theory on a fixed  torsional Newton-Cartan geometry --- a natural  generalization of Newton–Cartan geometry  that allows for torsion and correctly encodes the symmetries of non-relativistic field theories coupled to curved backgrounds 
\cite{Christensen:2013lma,Christensen:2013rfa}. The definition of the non-relativistic  stress-energy tensor complex in Lifshitz duality was proposed from the action principle based on vielbein fields
\cite{Ross:2009ar}. Important progress on Lifshitz holography can be found in e.g. \cite{Hartong:2014pma, Bergshoeff:2014uea,  Chemissany:2014xsa, Baggio:2011cp, Mann:2011hg, Kiritsis:2015doa, Gursoy:2016tgf}. 

We exploit Lifshitz holography to study the shear sector of the hydrodynamic response from Einstein-Maxwell dilaton theory. Using linear response theory, we systematically extract the low-energy, long-wavelength excitations and extract the structure of the hydrodynamic pole associated with transverse momentum fluctuations from quasinormal modes in the bulk. The consistent metric and gauge field  fluctuations in the bulk are dual to perturbations on the TNC boundary geometry. 
We employ a combination of analytical  matching method (near-horizon and far-region matching) and  numerical method to obtain the dispersion relation of the hydrodynamic pole in the shear channel. 
 
The organization of this paper is as follows. In Sec. \ref{s1}, we briefly introduce the EMD model, and its connection with the effective field theory defined in the Newton-Cartan geometry. Sec. \ref{s3} is the main part of this work. In Sec. \ref{fluctuation}, we do the transverse fluctuation to study the shear dynamics of momentum, and demonstrate the shear subdiffusion of momentum in Sec. \ref{sec:matching} by the matching method. In Sec. \ref{sec:num}, we perform the numerical computations of the lowest quasi-normal modes to confirm our analytical results and show the property of the subdiffusive constant. Finally, in Sec. \ref{sec:conclusion}, we conclude for this study and discuss some open questions for the future. 

\vspace{0.4cm}
\section{The holographic model and the generating functional}\label{s1}

We begin by outlining the bulk  Einstein-Maxwell-dilaton theory which  supports asymptotically Lifshitz black hole solutions. Then 
we discuss the generating functional on the  boundary geometry, which takes the form of a torsional Newton-Cartan spacetime. 
\subsection{The Einstein-Maxwell-dilaton model}
We start from the Einstein-Maxwell-dilaton theory \cite{Taylor:2008tg} that is described by the action in the bulk
\begin{equation}
 S=\frac{1}{16\pi G}\int d^{d+2} x \sqrt{-g} \,\bigg [  R - 2\Lambda  - \frac{1}{4} e^{\lambda \phi} F^2 - \frac{1}{2} (\nabla \phi)^2 \bigg]\,,
\end{equation}
where the cosmological constant is 
\begin{equation}
\Lambda = - \frac{1}{2} ( z + d )( z + d - 1)\,.
\end{equation}
The gauge field $A_a$ couples to a dilaton field $\phi$ with a special chosen of interaction where the field strength $F_{ab}=\partial_a A_b-\partial_b A_a$.\footnote{We use the index $a,b$ for the fields in the bulk, and the index $\mu,\nu$ for fields at the boundary. } As a result, it deforms the UV boundary to be asymptotic to a Lifshitzian spacetime, where the parameter
\bea
\lambda = \sqrt{ \frac{2d}{ z - 1}}\,
\eea
is a function of the Lifshitz exponent $z$ with $z>1$. It is obvious that with different choice of $\lambda$, we have different Lifshitz exponent. Here $d$ is the spatial dimension of the dual field theory, and we will specify to $d=2$ in the following  when we consider the hydrodynamics in the dual field theory. Extension to other dimensions is straightforward. 
We set $16\pi G=1$ without loss of generality. 

The equations of motion can be obtained after performing the variation, 
\begin{equation}
\begin{split}
R_{ab} - \frac{1}{2} e^{\lambda \phi} F_{a c} F_{b}^{~c} - \frac{1}{2} \partial_a\phi \partial_b\phi
&= \frac{1}{2} g_{ab} \left[R - 2\Lambda - \frac{1}{4} e^{\lambda \phi} F^2 - \frac{1}{2} (\nabla \phi)^2 \right]\,, \\
\nabla_a (e^{\lambda \phi} F^{ab}) &= 0\,, \\
\nabla^2\phi - \frac{\lambda}{4} e^{\lambda \phi} F^2 &= 0\,.  
\end{split}
\end{equation}
This model has a series of analytic solutions that dual to the finite temperature and finite (particle) density many-body systems, with the line element written as
\bea
\label{eq:geometry}
d s^2 = -r^{2z} f(r) d t^2 + \frac{dr^2}{r^2 f(r)} + r^2 d {\vec{x}_{d}}^2\,, 
\eea
where 
\bea
\begin{split}
f(r)=1 - \left(\frac{r_h}{r}\right)^{z+d}\,,\quad
A_t =\sqrt{ \frac{2(z-1)}{z+d} } \left(r^{z+d}- r_h^{z+d}\right)\,,
\end{split}
\eea
and 
\begin{equation}
\label{eq:background}
\begin{split}
e^{\lambda \phi} &= r^{-2d}\,. \\
\end{split}
\end{equation}
The temperature of the black hole is related to the horizon $r_h$ as \be T = \frac{z+d}{4 \pi}\,r_{h}^z\,.
\ee 

Remind that for a pure Lishitz spacetime with Lifshitz symmetry, the geometry should be invariant under the Lifshitz scaling transformation 
\bea\label{Lifshitz}
t\rightarrow c^z t\,,\quad
x^i\rightarrow c x^i\,,\quad
r\rightarrow c^{-1} r\,
\eea
with $z>1$ as the Lifshitz exponent. 
However, in this work, we consider the finite temperature and finite mass density systems, temperature and the matter fields together with the scalar field break the Lifshitz symmetry as the relevant deformations. 

The Lifshitz symmetry Eq.\eqref{Lifshitz} naturally breaks the Lorentz boost symmetry between the time and spatial directions. Consequently, the low-energy dynamics of the dual systems should not respect the relativistic symmetry anymore, and the description is replaced by the Galilean invariant non-relativistic field theories. 
Nevertheless, nonrelativistic  diffeomorphism invariance still constrains the low-energy dynamics for the non-relativistic quantum systems \cite{
Son:2013rqa, Geracie:2014nka}, especially in the strongly-coupled quantum systems with a holographic dual as our consideration. 

\subsection{The generating functional}
In a relativistic quantum field theory, the energy-momentum tensor $T^{\mu\nu}$ and the conserved $U(1)$ current  $J^\mu$ couple to the external background metric $\gamma_{\mu\nu}$ and gauge field $A_\mu$, respectively. In general, we can always include a scalar operator $\mathcal{O}$ that couples to external scalar source $\phi$. Therefore, the low-energy dynamics are captured by the generating functional
$W[\gamma_{\mu\nu}\,,A_\mu\,,\phi]$, with the variation
\be
\delta W=\int d^3x \sqrt{-\gamma}\big(
T^{\mu\nu}\delta\gamma_{\mu\nu}+J^\mu \delta A_\mu+\mathcal{O}\delta\phi
\big)\,.
\ee

For the non-relativistic theory, there is no underlying non-degenerate metric in the non-relativistic spacetime. 
It is described by 
a Newton-Cartan geometry \cite{Jensen:2014aia, Jensen:2014ama, Jensen:2014wha}, 
a $(d+1)$-dimensional manifold equiped with $(n_\mu, \sigma_{\mu\nu}, a_\mu)$, i.e. a no-where vanishing one-form $n_\mu$ which defines local time direction and a symmetric spatial tensor $\sigma_{\mu\nu}$ of rank-$d$ which gives the metric on spatial slices, and a U(1) connection $a_\mu$ whose field strength leads to a unique definition of the covariant derivative, satisfying 
\bea
\gamma_{\mu\nu}= n_\mu n_\nu+\sigma_{\mu\nu}\,,
\eea
with rank-$(d+1)$ and invertible $\gamma^{\mu\nu}$. 
Then a velocity field $v^\mu$ and an inverse metric $\sigma^{\mu\nu}$ can be straightforwardly defined via 
\be
v^\mu n_\mu=1\,,~~~ \sigma_{\mu\nu}
v^\mu=0\,,~~~
\sigma^{\mu\nu}
n_\nu=0\,,~~~
\sigma^{\mu\alpha}\sigma_{\nu\alpha}=\delta^\mu_\nu-v^\mu n_\nu\,.
\ee
The one form $n_\mu$ effectively define a local time direction and $\sigma_{\mu\nu}$ is a metric on spatial slices.  
Equivalently, $(v^\mu,\sigma^{\mu\nu})$ satisfy  
\be
v^\mu=\gamma^{\mu\nu} n_\nu\,,~~~ \sigma^{\mu\nu}=\gamma^{\mu\nu}-v^\mu v^\nu\,.
\ee
In torsionless Newton-Cartan  geometry the one form $n_\mu$ is exact, i.e. $dn=0$. Otherwise it is torsional Newton-Cartan  geometry with $dn\neq 0$ which is the case we focused on. 

Along with the reduction of the geometry, the energy-momentum tensor is decomposed to the energy density $E$, the energy flux $E^i$, the momentum density $P^i$, and the symmetric  stress tensor $\pi^{ij}$ with only the spatial components, as well a number current $J^\mu$. 
The generating functional is therefore expressed by the Newton-Cartan data 
$W[n_\mu, v^\mu, \sigma^{\mu\nu}, a_\mu, \phi_0]$, with the variation
\be
\label{eq:nr-sv}
\delta W=\int d^3x \sqrt{-\gamma}\left(
J^\mu \delta a_\mu
-P_\mu \delta {\bar v}^\mu
- E^\mu \delta n_\mu -\frac{1}{2}\pi_{\mu\nu}\delta {\bar {\sigma}}^{\mu\nu}
+\mathcal{O}\delta \phi_0\right)\,, 
\ee
 where $\delta {\bar v}^\mu$ is the free variation part of  $\delta v^\mu$ excluding terms involving $\delta n_\mu$. 
Imposing the $U(1)$ gauge invariance leads to the particle number conservation $\partial_t n + \partial_iJ^i=0$. In the special system we studied, it turns out $\delta a_y$ is not indepedent of $\delta n_y$ which means that we only have an independent $\delta a_t$ while there is no independent source for $J^i$ \cite{Taylor:2015glc}.   
In the absence of boost invariance, imposing the time and spatial translational symmetry, the energy and momentum are independently conserved (assuming vanishing scalar source) 
\bea
\partial_t E + \partial_i E^i=0\,,\quad
\partial_t P_j + \partial_i \pi^i_{~j}=0\,.
\eea



\section{Transverse fluctuation and the shear diffusion}
\label{s3}

In this section we perform  holographic calculations for the transverse fluctuations. First, we identify the correct gauge invariants with independent sources. Subsequently, we use the near-far matching method to derive the dispersion relation of the hydrodynamic diffusive  mode. We also  numerically compute the quasinormal modes of the system and compare with the analytical results.  

\subsection{The linear transverse  perturbations}\label{fluctuation}
To investigate the dynamics of the momentum, we switch on the linear fluctuations of the metric and matter fields in the shear channel. In general, the linear fluctuations of the metric, gauge field and the scalar field are the following
\begin{equation}
\begin{split}
\label{eq:pertur}
\delta g_{ab} &= \int \frac{d\omega dk}{(2\pi)^2} e^{-i\omega t + i k x} h_{ab}(r,\omega,k)\,, \\
\delta A_{a} &= \int \frac{d\omega dk}{(2\pi)^2} e^{-i\omega t + i k x} a_{a}(r,\omega,k)\,, \\
\delta \phi &= \int \frac{d\omega dk}{(2\pi)^2} e^{-i\omega t + i k x} \psi(r,\omega,k)\,, \\
\end{split}
\end{equation}
after the Fourier transformations. 

These fluctuations can be classified by parity under $y\to -y$. Consider the parity-odd fluctuations $\{h_{t y},h_{x y},a_{y}\}$ around the background (\ref{eq:background}) in the radial gauge $\delta g_{r i} = \delta a_r=0$, they follow the EoMs 
\begin{subequations}
\label{radialflctuations}
\begin{align}
h_{t y}''- \frac{z-1}{r} h_{t y}' + \frac{-k^2 + 2r^2(z-2)f}{r^4 f} h_{t y} - \frac{k \omega}{r^4 f} h_{x y} + e^{\lambda \phi} A_{t}' a_{y}' &= 0 \,,\\
h_{x y}'' + \frac{(z-1)f + r f'}{r f} h_{x y}'+ \frac{\omega^2 - 2r^{2z}f(z f + r f')} {r^{2(z+1)}f^2} h_{x y} + \frac{k \omega}{r^{2(z+1)}f^2} h_{t y} &= 0 \,,\\
a_{y}'' + \frac{r f'+f(1+z+r \lambda \phi')}{r f} a_y' + \frac{r^{2(1-z)}\omega^2 - k^2 f}{r^4 f^2} a_y + \frac{A_{t}'}{r^{2z}f}h_{t y}' - \frac{2A_{t}'}{r^{1+2z}f}h_{t y} &=0 \,,\\[9pt]
\omega r h_{t y}' - 2\omega h_{t y} + k r^{2z-1} f h_{x y}' - 2k r^{2(z-1)}f h_{x y} + \omega r e^{\lambda \phi} A_{t}' a_y &= 0 \,.
\end{align}
\end{subequations}
The EoMs of the parity-odd sector are decoupled from the parity-even sector including $\{h_{t t}, h_{t x}, h_{x x},h_{y y}, a_{t}, a_{x}, \delta\phi\}$. 

With the EoMs of the fluctuations, the standard approach to compute the quasi-normal modes of the system is to solve the ODEs and impose the infalling boundary conditions near the horizon and the vanishing Dirichilet boundary conditions at the UV Lifshitz boundary. As for the AdS scenarios, choosing the gauge invariant variables can make the computations simpler. The gauge invariant variables are combinations of the fluctuations that are invariant under diffeomorphism and gauge transformations.  
To be concrete, under an infinitesimal diffeomorphism generated by the vector field $\xi_{\nu}$ and an infinitesimal gauge transformation generated by scalar field $\lambda$, the metric and gauge field fluctuations transform as
\begin{equation}
\begin{split}
\delta g_{ab} &\to \delta g_{ab} - \nabla_{a}\xi_{b} - \nabla_{b}\xi_{a}\,,\\
\delta A_{a} &\to \delta A_{a} + \nabla_{a}\lambda - \xi^{b}\nabla_{b}A_{a} - A_{b}\nabla_{a}\xi^{b},\\
\delta \phi &\to \delta \phi - \xi^{a}\nabla_{a}\phi\,, \\
\end{split}
\end{equation}
which constrains the number of independent fields. Therefore, there are only two independent gauge invariant variables in the transverse channel, defined as 
\begin{equation}\label{gaugeinvariantvariables}
Z_1 \equiv \omega h_{x y} + k h_{t y}\,, \quad Z_2 \equiv a_y\,.
\end{equation}

The form of these two gauge invariants is a common choice in the AdS references or the special Lifshitz spacetime with $z=1$. 
However, this choice is problematic when we generalize to the Lifshitz spacetime with $z>1$, since the leading order solution of $Z_1$ and $Z_2$ are linearly dependent where
\begin{equation}
\begin{split}
Z_{1} &= r^{2z} \left( Z_{1}^{(0)} + Z_{1}^{(1)} r^{-z-2}+ ... \right)+r^2 Z_{3}^{(0)},\\
Z_{2} &= r^{z+2} \left( Z_{2}^{(0)} + Z_{2}^{(1)} r^{-z-2}+ ... \right),\\
\end{split}
\end{equation}
with
\be 
Z_{1}^{(0)} = - k \sqrt{\frac{z+2}{2(z-1)}} r^{z-2} Z_{2}^{(0)}\,. 
\ee
This is a special property in Lifshitz holography, which is different from the AdS cases. We will show that this is because the energy flux $E_i$ and the momentum density $P_i$ are independent in the non-relativistic system, while their dynamics are mixed in the fluctuation $h_{ty}$. 
Via the holographic dictionary, the leading order solutions $Z_1^{(0)}$ and $Z_2^{(0)}$ are the external sources in the field theory, while the sub-leading solutions $Z_1^{(1)}$ and $Z_2^{(1)}$ are the operators that couple to the external sources. 
In order to extract another independent source that is represented by $Z_3^{(0)}$, 
we make a linear combination of $Z_{1}$ and $Z_2$ to define a new gauge invariant
\be
\label{eq:gau-inv-z3}
Z_3=\omega h_{x y} + k h_{t y} + k\, \sqrt{\frac{z+2}{2(z-1)}}\, r^{z-2}\, a_y\,,
\ee
with the asymptotic behaviors of $Z_3$ as
\begin{equation}
\begin{split}
Z_{3} &= r^2 Z_{3}^{(0)}+... \,.\\
\end{split}
\end{equation}
As a result, we remove the leading order contribution of $Z_1^{(0)}$ and now the source $Z_3^{(0)}$ is dominant as the source in $Z_3$. 

Now we have a consistent choice of gauge invariant variables, i.e. $\{Z_2\,, Z_3\}$ for further studies of quasi-normal modes as well as the hydrodynamics in this system. However, before the solving the ODEs and read out the hydrodynamic poles, it is necessary to explain the physical meaning of the variables and make a connection with the non-relativistic hydrodynamics.

In the non-relativistic systems where the boost symmetry is broken, the momentum density $P_y$ is not equal to the energy flux $E^y$ and they couple to independent source $\hat{e}^{(y)}_{~~t}$ and $\hat{e}^{(t)}_{~~y}$, as shown in \eqref{eq:nr-sv}. 
This can be made clear from frame fields defined from Lifshitz geometry \cite{Ross:2009ar, Ross:2011gu}. 
We use $e^{(0)}, e^{(i)}$ to construct the frame fields in the bulk, while we use $\hat{e}^{(0)}, \hat{e}^{(i)}$ with a hat to construct the frame fields in the boundary field theory. 
The bulk metric can be expressed as  $ds^2=g_{ab}dx^a dx^b\equiv(e^{(r)})^2+\eta_{\mu\nu}e^{(\mu)}e^{(\nu)}$ in terms of the bulk frame fields $e^{(\mu)}$. Then the frame field $\hat{e}^{\mu}$ at the boundary can be determined from 
\bea
e^{(0)}= r^z\sqrt{f}\hat{e}^{(0)}\,,~~~~e^{(i)}= r\hat{e}^{(i)}\,,
\eea
in the limit $r\to\infty$.

For the background bulk solution Eq. \eqref{eq:geometry} without any fluctuations, the frame fields are defined as
\bea
e^{(0)}=e^{(0)}_{~~\mu}dx^\mu=r^z\sqrt{f}dt\,,~~~e^{(i)}=e^{(i)}_{~~\mu} dx^{\mu}=r\left(dx+dy\right)
\eea
which defines the vielbeins
\bea
e^{(0)}_{~~t}=r^z\sqrt{f}\,,~~~e^{(0)}_{~~x}=e^{(0)}_{~~y}=0\,,~~~e^{(i)}_{~~t}=0\,,~~~e^{(i)}_{~~j}=r\delta_{ij}\,.
\eea

When we switch on the transverse sector fluctuations $h_{ty},h_{xy},a_y$, we follow \cite{Ross:2009ar} to reparametrize the fluctuations as\footnote{The fluid/gravity duality for five dimensional  asymptotically Lifshitz black hole was  explored in \cite{Kiritsis:2015doa}, where a Galilean boost was applied to construct perturbative solution. In contrast to the  fluctuations considered in the present work, the analysis in \cite{Kiritsis:2015doa} did not include  nonzero and non-homogeneous $e^{(t)}_{~~y}$ that sources the energy flux $E^y$. 
These components, however, plays a crucial role in generating the full TNC structure of the dual boundary field theory \cite{Ross:2009ar, Ross:2011gu,
Christensen:2013rfa, 
Christensen:2013lma}. Consequently, the study \cite{Kiritsis:2015doa} effectively captured only the standard (torsion-free) Newton–Cartan geometry. It would be interesting to perform a proper fluid/gravity duality for our case and make a  comparison to those derived  here.}
\bea
h_{ty}=-r^{2z}v_{1}+r^2v_{2}\,,~~~h_{xy}=r^2 v_3\,,~~~a_y=\sqrt{\frac{2(z-1)}{z+2}} r^{z+2}v_{1}\,. 
\eea
In Lifshitz holography, the shear stress is encoded in the gravitational fluctuation $h_{xy}$ as for the AdS scenarios. However, the independent external sources time shift and spatial shift are mixed in the gravitational fluctuation $h_{ty}$ and the different conformal dimensions add further complications. 
Furthermore, note that the variation of $a_\mu$ is linked to $h_{ty}$. This implies that the gauge transformation of the transverse gauge field is not independent of the time shift, which constitutes a constraint on the torsional Newton-Cartan background.    
This special feature puts a constraint on $J^y$ and $E^y$ in 
\eqref{eq:nr-sv}.

It is convenient and equivalent to make the following variation in the frame fields as
\be
\begin{split}
\label{eq:e0i}
e^{(0)}&=r^z\hat{e}^{(0)}= r^z \left[dt+v_{1}dy\right] \,,\\
e^{(i)}&=r\hat{e}^{(i)}=r\left[ v_{2} d t +\left(\delta^i_{~j}+v_3\bar{\delta}^i_{~j}\right)d x^j\right]\,,
\end{split}
\ee
where $v_1,v_2,v_3$ are functions of $r,t,x$ and $\bar{\delta}^i_{~j}\equiv 1-\delta^i_{~j}$. When $r\to\infty$, the leading order of $v_i~ (i=1,2,3)$ are constants of $\mathcal{O}(r^0)$.

The asymptotic solutions near the boundary for $1<z<2$  are 
\bea
\label{eq:vi-bnd-exp}
\begin{split}
v_{1}(r,t,x)&=s_{1}+\frac{c_{1}}{r^{3z}}+\frac{d_{1}}{r^{z+2}}+\cdots\,,\\
v_{2}(r,t,x)&=s_{2}+\frac{c_{2}}{r^{4-z}}+\frac{d_{2}}{r^{z+2}}+\cdots\,,\\
v_3(r,t,x)&=s_3+\frac{d_3}{r^{z+2}}+\cdots\,.
\end{split}
\eea
It is necessary to make some explanations here
\begin{itemize}
\item The expansions in Eq.\eqref{eq:vi-bnd-exp} are valid for $1<z<2$, and  there are more restrictions for $z\geq 2$ \cite{Ross:2011gu}. In this work, we focus on the non-relativistic systems with $1<z<2$.
\item All the parameters $s_{1,2,3}$, $c_{1,2}$ and $d_{1,2,3}$ are functions of $r,t,x$. $s_{1,2,3}$ are the sources, $c_{1,2}$ are the two independent responses, and $d_{1,2,3}$ depend on both $s_i$ and $c_i$. One of these three sources can be set to zero by residual gauge transformation. This is consistent with the fact that there are only two independent gauge invariants. 
\item Different from AdS cases, the full UV asymptotic expansions can be very complicated since $\pm z$ is involved in the power expansions. We omit the expressions for these correction in ellipsis. 
\item The first vielbein field $v_1(r,x^\mu)$ with the boundary condition $v_1(\infty,x^\mu)=s_1\equiv \hat{e}^{(t)}_{~y}$ corresponds to the time shift transformation $t\rightarrow t+\hat{e}^{(t)}_{~y} y$ that induces the energy flux $E^y$ that is proportional to the bulk data $c_1$. 
\item The second vielbein field $v_2(r,x^\mu)$ with the boundary condition $v_2(\infty,x^\mu)=s_2 \equiv \hat{e}^{(y)}_{~t}$ represents the spatial shift $y\rightarrow y + \hat{e}^{(y)}_{~t}t$ that induces the momentum density $P_y$ that is proportional to the bulk data $c_2$. 
\item The third vielbein field $v_3(r,x^\mu)$ with the boundary condition $v_3(\infty,x^\mu)=s_3\equiv \hat{e}^{(y)}_{~x}$ induces the symmetric shear stress tensor $\pi^x_{~y}$ that is also proportional to the bulk data $c_2$, as a result of the conservation law of momentum density $P_y$. 
\item Finally, in the power expansion of $v_1$, the response $c_1$ induced from $s_1$ can be irrelevant compared to $d_1$ induced from $s_2$. We can use the analytic solutions to identify the correct responses, as shown in Eq.\eqref{eq:solva0} and Eq.\eqref{eq:solvb0}.  
\end{itemize}

Performing the holographic renormalization \cite{Ross:2009ar, Ross:2011gu} for the asymptotic Lifshitz bulk \eqref{eq:geometry}, we obtain the effective action at the linear order
\bea
\delta S=\int d^3x\left[-E^y\delta\hat{e}^{(0)}_{~~y}+P_y \delta\hat{e}^{(y)}_{~~t}+\pi^x_{~y}\delta\hat{e}^{(y)}_{~~x} \right]\,,
\eea
which serves as the Lifshitzian holographic dictionary that defines the momentum density $P_y$, the energy flux $E^y$ and the momentum flux $\pi^x_{~y}$
\bea
\begin{split}
P_y&=r^{z+2}\left(-rv_1'+r^{3-2z}v_2'\right)\Big|_{r\rightarrow \infty}\,,\\
E^y&=\frac{z-4}{z+2}r^{1+3z}v_1'+r^{3+z}v_2'\Big|_{r\rightarrow \infty}\,,\\
\pi^x_{~y}&=-r^{3+z}fv_3'\Big|_{r\rightarrow \infty}\,.
\end{split}
\eea
Comparing the momentum density and momentum flux  with the first-order constraint equation 
\bea
r\omega v_1'-r^{3-2z}\omega v_2'=krfv_3'\,,
\eea
we obtain the conservation equation for the momentum density as
\bea
\partial_t P_y+\partial_x \pi^x_{~y}=0\,,
\eea
where $\partial_t\rightarrow -i\omega$ and $\partial_x\rightarrow ik$.

In terms of the vielbein fields $v_{1,2,3}$, the gauge  invariants $\{Z_2, Z_3\}$ defined in \eqref{gaugeinvariantvariables} and \eqref{eq:gau-inv-z3} can be written as
\bea
Z_2=\alpha r^{z+2} v_1\,, \quad\quad 
Z_3=r^2\left(kv_2+\omega v_3\right)\,.
\eea
For latter convenience in the analytical computations, we redefine a new set of gauge invariants 
\be
V_a \equiv k v_1\,,~~~ V_b=kv_2+\omega v_3\,
\ee
where both $V_a$ and $V_b$ approaches constants asymptotic to the UV boundary at the leading order from  \eqref{eq:vi-bnd-exp},
\bea\label{eq:asy-vab}
V_a=s_a+\frac{c_{a}}{r^{3z}}+\frac{d_{a}}{r^{z+2}}+\cdots\,,~~~V_b=s_b+\frac{c_{b}}{r^{4-z}}+\frac{d_{b}}{r^{z+2}}+\cdots\,.
\eea
with $s_a=ks_1\,, c_a=kc_1\,,d_a=kd_1\,,s_b=ks_2+\omega s_3\,,c_b=kc_2\,,d_b=kd_2+\omega d_3$. 
 Note that $d_{a,b}$ are functions of $s_{a,b}$ and $c_{a,b}$.  
The EoMs of $V_{a,b}$
can be written in the following form
\begin{equation}
\label{EomTotalD}
\begin{split}
0&=\frac{d}{dr}\Big[r^{1+z}f\Big(\left(z-4\right)r^{2z}f^2\Big(\frac{V_a}{f}\Big)'+\left(z+2\right)r^2V_b'\Big)\Big]
\,\\&~~~~~~~~~~+\frac{r^2\omega^2-k^2r^{2z}f}{r^{z+3}f}\Big(\left(z-4\right)r^{2z}fV_a+\left(z+2\right)r^2V_b\Big)\,,\\\vspace{.1cm}
0&=\frac{d}{dr}\Big[\frac{r^{z+3}f}{r^2\omega^2-k^2r^{2z}f}\left(r^{2z}V_a'-r^2V_b'\right)\Big]+\frac{r^{2z}V_a-r^2V_b}{r^{z+1}f}\,.
\end{split}
\end{equation}
Eq. \eqref{EomTotalD} is the starting point in the following computations. In Sec. \ref{sec:matching}, we solve Eq. \eqref{EomTotalD} analytically and iteratively via the matching method to obtain the hydrodynamic dispersions and the subdiffusive constant. In Sec. \ref{sec:num}, we perform the numerical computations to obtain the gapless quasi-normal modes, and compare the results with our analysis.

\subsection{Holographic  calculation from matching method}
\label{sec:matching}
To obtain the retarded Green's function for the non-relativistic hydrodynamics, we divide the spacetime outside the horizon into two regions,  namely, the 
(near horizon) inner region 
\be\label{eq:nhregion}
\frac{r-r_h}{r_h}\ll 1\,,
\ee
and the (far from horizon) outer region \be
\frac{\omega^2}{ r^{2z} f^2 }\ll 1\,,~~~\frac{k^2}{ r^2 f } \ll 1\,,
\ee 
and solve the
equations \eqref{EomTotalD} in each region separately. 
After obtaining the solutions, we perform the matching between them in the overlap region
\be\label{eq:matchingregion}
\text{max}\bigg(\frac{\omega}{2\pi T}\,, \frac{k}{(2\pi T)^{1/z}}\bigg)\,\ll\frac{r-r_h}{r_h}\ll 1\,.
\ee

\subsubsection{The inner solutions} 
In the near region  \eqref{eq:nhregion}, 
Eq. \eqref{EomTotalD} becomes
\bea
V_a''+V_a'\left[\frac{1}{r-r_h}+...\right]+V_a\left[\frac{\omega^2}{(z+2)^2r_h^{2z}(r-r_h)^2}+...\right]+V_b'\left[...\right]&=0\,,\\
V_b''+V_b'\left[\frac{1}{r-r_h}+...\right]+V_b\left[\frac{\omega^2}{(z+2)^2r_h^{2z}(r-r_h)^2}+...\right]+V_a'\left[...\right]+V_a\left[
...\right]&=0\,,
\eea
where the ellipses represent the higher-order terms in the near-horizon expansion. The EoMs of $V_a$ and $V_b$ decouples at the horizon, which admits the infalling and outgoing solutions for $V_{a,b}$
\bea
V_{a,b}
=v^+_{a,b}\,\text{exp}
\Big[
 \frac{i \omega}{ 4 \pi T } \text{ln}\left(r-r_h\right) \Big]+ v^-_{a,b}\,\text{exp}
\Big[
- \frac{i \omega}{ 4 \pi T } \text{ln}\left(r-r_h\right) \Big]\,.
\eea
We impose the infalling boundary conditions near the Lifshitz black hole to obtain the retarded Green's function in non-relativistic field theory,  
and obtain
\bea
\label{eq:nearexp}
V_{a,b}^{\text{inner}} 
=\Big(\sum_{n=0} \nu_{a,b}^{(n)}\left(r-r_h\right)^n \Big)\,\text{exp}
\Big[
- \frac{i \omega}{ 4 \pi T } \text{ln}\left(r-r_h\right) \Big]\,,
\eea
where $\nu_{a,b}^{(0)}$ are independent variables, and $\nu_{a,b}^{(n)}(n>0)$ are functions of $\nu_{a,b}^{(0)}$ and can be solved order by order. As long as $\omega$ and $k$ are sufficiently small, these series solutions are valid from the horizon to some UV cutoff, which defines the inner region. 

\vspace{0.4cm}
\subsubsection{The outer solutions} 
We can solve the equations iteratively in the outer region of the bulk, where the frequency $\omega$ and momentum $k$ are small quantities compared to the radial coordinate $r$
\begin{equation}\label{eq:omegakepsilon}
\frac{\omega^2}{ r^{2z} f^2 } \ll 1\,,~~~~\frac{k^2}{ r^2 f } \ll 1\,.
\end{equation}
One can confirm that for finite $\omega$ and $k$, $\frac{\omega^2}{ r^{2z} f^2 }$ and $\frac{k^2}{ r^2 f }$  are negligible in the ODEs as $r$ approaches the UV boundary. As $r$ approaches the IR horizon, these two terms can still remain small in the hydrodynamic limit, which extends the valid range of the far-region that is possible to share overlap with the near-region. 
Notice that the contributions containing these infinitesimal small  quantities are additive and linearly independent in Eq.  \eqref{EomTotalD}. This indicates that it is not necessary to assume the order relations between these two terms. 
On the contrary, the scaling relation between $\omega$ and $k$ can be determined when we solve the equations in the full bulk. In other words, the pole condition can be extracted by imposing a specific relation on their respective orders within the matching region.\footnote{ This differs from the assumption used in  \cite{Mukherjee:2017ynv} where it is assumed that  the frequency $\omega$ and momentum $k$ are of same order from the beginning.
Together with an additional assumption that the two  variables are of same order $\mathcal{H}\sim  a_y$, whereas the  correct relation should be $\mathcal{H}\sim q a_y$, that approach \cite{Mukherjee:2017ynv} leads to a different structure in the leading-order compared to  \eqref{eq:solvb0}.  
In our calculation, the  orders of $\omega$ and $k$ are initially treated as  independent, and their relation is determined through the matching approach in the hydrodynamic limit. } 

The existence of the two small quantities \eqref{eq:omegakepsilon} makes it possible to construct the solutions iteratively in the far-region. 
Taking the first equation in Eq. \eqref{EomTotalD} as an example, in this region the coefficients of the non-derivative terms that are proportional to $V_a$ and $V_b$ are small quantities 
and therefore the non-derivative terms in the second line are negligible compared to the total derivative terms in the first line. We use similar observations to simplify Eq. \eqref{EomTotalD} into two coupled ODEs involving only total derivatives. 
\begin{equation}
\begin{split}
0&=\frac{d}{dr}\Big[r^{1+z}f\Big(\left(z-4\right)r^{2z}f^2\Big(\frac{V_a^{(0)}}{f}\Big)'+\left(z+2\right)r^2V_b^{(0)'}\Big)\Big]\,,\\
0&=\frac{d}{dr}\Big[\frac{r^{z+3}f}{r^2\omega^2-k^2r^{2z}f}\left(r^{2z}V_a^{(0)'}-r^2V_b^{(0)'}\right)\Big]\,.
\end{split}
\end{equation}
Here we have used the label $``(0)"$ to represent the solutions to the zeroth order. 
After integration once over the radial direction $r$, we obtain
\begin{equation}
\begin{split}
\left(z-4\right)r^{2z}f^2\Big(\frac{V_a^{(0)}}{f}\Big)'+\left(z+2\right)r^2V_b^{(0)'}&=\frac{(\pi_a-\pi_b)\omega^2}{r^{1+z}f}\,,\\
\frac{r^{z+3}f}{r^2\omega^2-k^2r^{2z}f}\left(r^{2z}V_a^{(0)'}-r^2V_b^{(0)'}\right)&=\frac{\pi_b}{z+2}\,,
\end{split}
\end{equation}
where $\pi_a, \pi_b$ are introduced as integration constants that correspond to operators in the effective action after some linear combinations, and they depend on frequency $\omega$ and momentum $k$.

It is simple to decouple $V_a^{(0)}$ to obtain a first-order ODE of $V_a^{(0)}$, and the solution of $V_a^{(0)}$ is expressed as an integral
\bea
\label{eq:solva0}
\begin{split}
V_a^{(0)}(r)=&\left(z+2+(z-4)f\right)\,\Big(\sigma_a+
\int_{\infty}^r d\rho\frac{\pi_a\rho^2\omega^2-\pi_bk^2\rho^{2z}f}{\rho^{3z+3}f\left(z+2+(z-4)f\right)^2}
\Big)\,,
\end{split}
\eea
where $\sigma_a$ is the integration constant that corresponds to the external source in the field theory. 
After obtaining the solution of $V_a^{(0)}$, we can obtain $V_b^{(0)}$ as 
\bea
\label{eq:solvb0}
V_b^{(0)}(r)=\sigma_b+\int_\infty^r d\rho
\Big[\pi_b\frac{k^2\rho^{2z}f-\rho^2\omega^2}{\rho^{z+5}(z+2)f}+\frac{V_a^{(0)'}}{\rho^{2-2z}}\Big]\,,
\eea
where $\sigma_b$ is the integration constant that corresponds to another external source. Here the asymptotic expansion of the above form is consistent with the expansion in \eqref{eq:asy-vab} up to a constant coefficient in front of $\sigma_a$. The UV expansions of $V_{a,b}^{(0)}$ are consistent with Eq. \eqref{eq:asy-vab}. One can check the integrals are convergent in the UV for $z<4$. However, we will encounter divergence when we consider higher order corrections, as discussed in Eq. \eqref{eq:solvc} for $z\geq 2$. 

Before our calculations of the ODEs to the next order, it is an appropriate time to make a comparison between the non-relativistic hydrodynamics and the relativistic hydrodynamics. In relativistic hydrodynamics, calculations to this order are sufficient to determine the hydrodynamic pole structure \cite{Davison:2013bxa}. However, as demonstrated in Subsection  \ref{subsec:matching}, a higher-order analysis is crucial for obtaining the correct pole structure in our case. This parallels the findings of recent studies on low-temperature matching for RN black holes  \cite{Gouteraux:2025kta}. 

For latter convenience in calculations, we assume
\begin{equation}
\label{eq:far}
V_{a,b} = V_{a,b}^{(0)} + V_{a,b}^{(c)} + \cdots,
\end{equation}
where $V_{a,b}^{(c)}$  represent the next-to-leading order corrections in terms of parameters \eqref{eq:omegakepsilon} to $V_{a,b}^{(0)}$. The dots represent higher order of $\epsilon$ corrections that are not important in this work.  $V_{a}^{(0)}$ and $V_{b}^{(0)}$ play the role of the source terms in the inhomogeneous ODEs of $V_{a,b}^{(c)}$ as follows
\begin{equation}
\begin{split}
r^{1+z}f\Big(\left(z-4\right)r^{2z}f^2\Big(\frac{V_a^{(c)}}{f}\Big)'+\left(z+2\right)r^2V_b^{(c)'}\Big)&=-h(r)\,,\\
\frac{r^{z+3}f}{r^2\omega^2-k^2r^{2z}f}\left(r^{2z}V_a^{(c)'}-r^2V_b^{(c)'}\right)&=-g(r)\,,
\end{split}
\end{equation}
where
\begin{equation}\label{g(r),h(r)}
\begin{split}
h(r) \equiv& \int_{r_h}^r dl \frac{l^2\omega^2-k^2l^{2z}f}{l^{z+3}f}\Big[\left(z-4\right)l^{2z}fV_a^{(0)}+\left(z+2\right)l^2V_b^{(0)}\Big] \,,\\
g(r) \equiv&  \int_{r_h}^r dl \frac{l^{2z}V_a^{(0)}-l^2V_b^{(0)}}{l^{z+1}f}\,.
\end{split}
\end{equation}
The lower bound of integration
is chosen so that the subleading solution remains  regular in the matching region. 

The solutions of $V_{a,b}^{(c)}$ contain the homogeneous part and the inhomogeneous part. The homogeneous part represent the higher order corrections in $\omega$ and $k$ to the integration constants $\sigma_{a,b}$ and $\pi_{a,b}$, and can be absorbed into $V_{a,b}^{(0)}$. As a result, we can expand $\sigma_{a,b}$ and $\pi_{a,b}$ order by order in $\omega$ and $k$. The inhomogeneous part of the correction order solution can be expressed as, 
\begin{equation}
\label{eq:solvc}
\begin{split}
V_{a}^{(c)}(r) = &  
\left(z+2+(z-4)f\right)\,\int_\infty^r d\rho
\frac{\left(z+2\right)\left(k^2\rho^{2z}f-\rho^2\omega^2\right)g(\rho)-\rho^2h(\rho)}{\rho^{3z+3}f\left(z+2+(z-4)f\right)^2}\,,\\
V_b^{(c)}(r)=&
\int_\infty^r d\rho\Big[\frac{\rho^2\omega^2-k^2\rho^{2z}f}{\rho^{z+5}f}g(\rho)+\frac{V_a^{(c)'}}{\rho^{2-2z}}\Big]\,.
\end{split}
\end{equation}
These integrals contribute to the corrections terms in the asymptotic UV expansions. They are convergent only if $1<z<2$. For $z\geq 2$, these integrals can contribute to positive power expansion in $r$ at UV, which is higher than the constant sources. Therefore, there are more restrictions for $z\geq 2$ \cite{Ross:2011gu}, and we leave the discussion on this case for the future. 

 Collecting \eqref{eq:solva0}, \eqref{eq:solvb0} and \eqref{eq:solvc} for \eqref{eq:far}, we obtain the boundary values of $V_a$ and $V_b$ with 
\be
\label{eq:nearbnd}
\begin{split}
V_a^{(s)}&=(2z-2)\sigma_a\,,~~~~
V_b^{(s)}=\sigma_b\,,~~~~
V_a^{(r)}\propto \frac{\omega^2}{2z-2}\pi_a\,,~~~~
V_b^{(r)}\propto \frac{k^2}{z+2}\pi_b\,.\\
\end{split}
\ee
Note that $V_{a,b}^{(s)}$ and $V_{a,b}^{(r)}$ plays the role of source and VEV respectively, with the quantities in the expansion of \eqref{eq:asy-vab} as $\sigma_a\propto s_a,\,\sigma_b\propto s_b$ and $\pi_a\propto c_a,\,\pi_b\propto c_b$.  

\subsubsection{Matching and dispersion relation}
\label{subsec:matching}

To manipulate the matching method, we expand the the far-region solutions in the inner region \eqref{eq:matchingregion} and compare with the inner solutions, i.e.  
\bea
\label{eq:nearexpan}
V_{a,b}^{\text{inner}} = \nu_{a,b}^{(0)} \Big( 1 - \frac{i \omega}{ 4 \pi T } \text{ln}\left(r-r_h\right) + \cdots \Big)\,.
\eea

To zeroth order solutions of $V_a^{(0)}$ and $V_b^{(0)}$, in the overlap region 
\bea
\begin{split}
V_a^{(0)}(r)=&~\frac{\pi_a\omega^2}{r_h^{3z}(z+2)^2}\text{ln}(r-r_h)+\,(z+2)\Big(\sigma_a+\frac{\pi_br_h^{-z-2}k^2}{2(z+2)^2(z-1)}\Big)+\mathcal{O}(\pi_a\omega^2)+\mathcal{O}(r-r_h)\,,\\
V_b^{(0)}(r)=&~\frac{(\pi_a-\pi_b)\omega^2}{r_h^{z+2}(z+2)^2}\text{ln}(r-r_h)+\,
\sigma_b+(z+2)r_h^{2z-2}\Big(\sigma_a+\frac{\pi_br_h^{-z-2}k^2}{2(z+2)^2(z-1)}\Big)\\
&~~~~+\mathcal{O}(\pi_a\omega^2,\pi_b\omega^2)+\mathcal{O}(r-r_h)\,,
\end{split}
\label{eq:innersol2}
\eea
where we have neglected terms of order $\omega^2$ in the coefficients of $(r-r_h)^0$ terms. The above expression should match with \eqref{eq:nearexpan}, i.e. one could get relations between the coefficients in front of $\ln(r-r_h)$ and $(r-r_h)^0$ terms, i.e. 
\bea
\begin{split}
\frac{i\pi_a\omega}{ r_h^{2z}(z+2)}&=(z+2)\Big(\sigma_a+\frac{\pi_br_h^{-z-2}k^2}{2(z+2)^2(z-1)}\Big)+\mathcal{O}(\pi_a\omega^2,\pi_b\omega^2)\,,\\
\frac{i(\pi_a-\pi_b)\omega}{ r_h^{2}(z+2)}&=\sigma_b+(z+2)r_h^{z-2}\Big(\sigma_a+\frac{\pi_br_h^{-z-2}k^2}{2(z+2)^2(z-1)}\Big) +\mathcal{O}(\pi_a\omega^2,\pi_b\omega^2)\,.
\end{split}
\eea
 From these relations, neglecting terms of order $\mathcal{O}(\pi_a\omega^2, \pi_b\omega^2)$ one could express $\pi_{a,b}$ as functions of $\sigma_{a,b}$ with\footnote{Note that this is the crucial difference compared with the shear diffusion in the relativistic hydrodynamics \cite{Davison:2013bxa}, 
 where matching the zeroth order far-region solutions with the near-region solutions leads to the momentum diffusion. }  
\bea
\begin{split}
\pi_a&=-i\frac{(z+2)^2}{\omega}r_h^{2z}\sigma_a+\frac{(z+2)k^2}{2(z-1)\omega^2}r_h^{z}\sigma_b\,,\\
\pi_b&=i\frac{(z+2)}{\omega}r_h^{2}\sigma_b\,.
\end{split}
\label{eq:matching-leading}
\eea
At the leading matched order, the dispersion relation has only one simple  pole at $\omega=0$, showing an almost  flat dispersion. 

This feature arises from the following fact due to the relation 
\bea
V_b^{(0)}=r_h^{2z-2}V_a^{(0)}-\frac{\pi_b\omega^2}{r_h^{z+2}(z+2)^2}\text{ln}(r-r_h)+\sigma_b\,. 
\eea
 This relation essentially gives the second relation in \eqref{eq:matching-leading}. 
Since the regular solutions lead to the operators/responses in the field theory, as shown latter, if we further do hydrodynamic expansions for the two combined operators based on $\pi_a$ and $\pi_b$, only one of the two combined operators involves the spatial derivatives to $k^2$ order. In other words, in the operator resulting from $r^{2z}V_a-r^2V_b$, the $k^2$ order contributions is absent, and we have to expand it to higher order,  i.e. $k^4$ from $r^{2z}V_a^{(c)}-r^2V_b^{(c)}$. 

The $k^4$ order contributions to the far-region solutions \eqref{eq:solvc}  are
\bea
\begin{split}\label{correction}
V_a^{(c)}(r)&=\pi_bk^4\Big(z+2+\left(z-4\right)f\Big)\Big(S(r)-S(\infty)\Big)+\mathcal{O}(\pi_a\omega^2,\pi_b\omega^2)\,,\\
V_b^{(c)}(r)&=\int_{\infty}^rd\rho\frac{V_a^{(c)'}}{\rho^{2-2z}}+\mathcal{O}(\pi_a\omega^2,\pi_b\omega^2)\,,
\end{split}
\eea
where $S(r)$ is defined as 
\bea
\label{eq:S(r)}
\begin{split}
S(r)=&~\Big[12  r_h^{10} (z-6) (z-2) (z-1) (z+2)^2r^2\Big]^{-1}\\[5pt]
&\bigg(\frac{12r_h^2(z-1)r^8-3(z-4)^2r_h^{10}}{\left(z-4\right)\left(z+2+\left(z-4\right)f\right)r^{z+2}}-3r_h^{8-z} \, _2F_1\left[1,\frac{-2}{z+2};\frac{z}{z+2};\tilde{r}^{z+2}\right]\\[5pt]
&-\frac{r^8}{r_h^z} \, _2F_1\left[1,\frac{6}{z+2};\frac{z+8}{z+2};\tilde{r}^{z+2}\right]\bigg)
\end{split}
\eea
that involves the hypergeometric functions and $\tilde{r}\equiv r/r_h$. With the solutions in Eq. \eqref{correction}, we compute the $k^4$ contribution to $V_b-r^{2z-2}V_a$ as
\bea
\begin{split}
V_b^{(c)}-r^{2z-2}V_a^{(c)}
=&~
\left(\rho^{2z-2}V_a^{(c)}\Big|^r_\infty\right)-\int_\infty^rd\rho \left(\rho^{2z-2}\right)'V_a^{(c)}-r^{2z-2}V_a^{(c)}\\[5pt]
=&-\int_\infty^rd\rho \left(\rho^{2z-2}\right)'V_a^{(c)}+\mathcal{O}(\pi_a\omega^2,\pi_b\omega^2)+\text{UV divergence}\,.
\end{split}
\eea
Here the ``UV divergence'' is due to the additional factor $r^{2z-2}$ on the left side  and encodes only the information of the external source, which is irrelevant when we match the far-region and near-region solutions in the matching region. 

With both the inner  and outer solution \eqref{eq:innersol2},  \eqref{correction} in the matching region, we can match them
\bea
\begin{split}
\frac{i\pi_a\omega}{ r_h^{2z}(z+2)}&=(z+2)\Big(\sigma_a+\frac{\pi_br_h^{-z-2}k^2}{2(z+2)^2(z-1)}\Big)+m_a k^4+\mathcal{O}(\pi_a\omega^2,\pi_b\omega^2)\,,\\
\frac{i(\pi_a-\pi_b)\omega}{ r_h^{2}(z+2)}&=\sigma_b+(z+2)r_h^{z-2}\Big(\sigma_a+\frac{\pi_br_h^{-z-2}k^2}{2(z+2)^2(z-1)}\Big)+m_b k^4 +\mathcal{O}(\pi_a\omega^2,\pi_b\omega^2)\,.
\end{split}
\eea
where 
\bea
\begin{split}
m_b-r_h^{2z-2}m_a&\equiv-\pi_b\,\mathcal{I}\\
&=-\pi_b\int_\infty^{r_h}d\rho(2z-2)\left(z+2+\left(z-4\right)f\right)\rho^{2z-3}\Big[S(\rho)-S(\infty)\Big]\,.
\end{split}
\eea
Therefore we obtain
\be
\begin{split}
\pi_a&= -i\frac{(z+2)^2}{\omega}r_h^{2z}\sigma_a-i\frac{(z+2)k^2}{2\omega(z-1)(-i\omega+(z+2)r_h^2\,\mathcal{I}\,k^4)}r_h^z\sigma_b\,,\\
\pi_b&=\frac{(z+2)r_h^{2}}{-i\omega+(z+2)r_h^{2}\,\mathcal{I}\,k^4}\sigma_b\,,\\
\end{split}
\ee
If one computes the corrections to next order, specifically terms of order  $\mathcal{O}(\pi_a k^4\omega^2, \pi_b k^4\omega^2, k^6)$, these will contribute an 
$\mathcal{O}(k^6)$ term to the denominator in the equations above.
From \eqref{eq:nearbnd}, 
we can read out a subdiffusive hydrodynamic mode in the correlator $\langle \pi_b \pi_b\rangle$, with the dispersion
\be
\label{eq:dis}
\omega=-i D_4 k^4+\mathcal{O}(k^6)\,,
\ee
where the leading order starts from $k^4$, $D_4$ is the subdiffusive constant
\bea
\begin{split}
D_4=&\left(z+2\right)r_h^2\,\int_\infty^{r_h}d\rho(2z-2)\left(z+2+(z-4)f\right)\rho^{2z-3}\Big[S(\rho)-S(\infty)\Big]\,,
\end{split}
\label{eq:D4}
\eea
where the function $S$ is defined in \eqref{eq:S(r)}. 

\vspace{0.4cm}
\subsection{Numerical verification}\label{sec:num}

\begin{figure}[h!]
\begin{center}
  \includegraphics[width=0.6\textwidth]{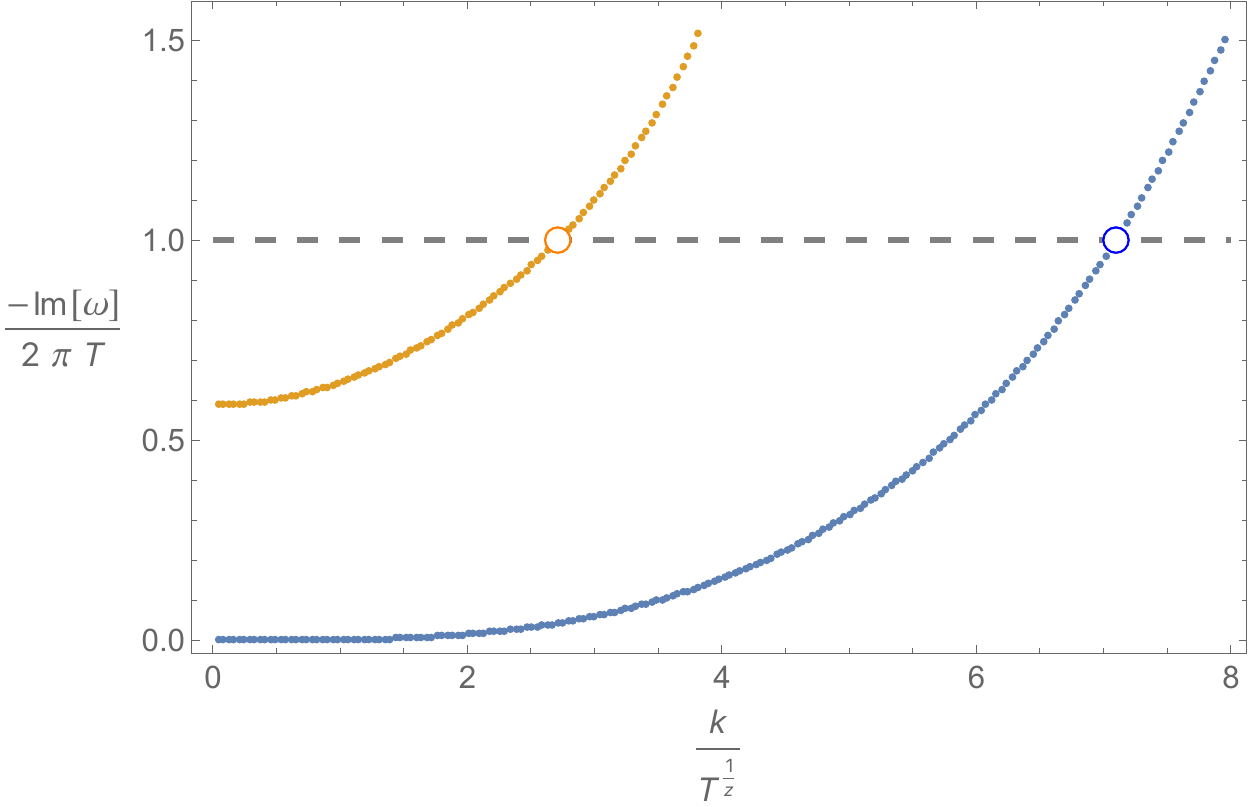}
  \end{center}
  \caption{\small Imaginary parts of frequencies of the hydrodynamic modes ({\em blue dots}) and non-hydrodynamic modes ({\em orange dots}) for $z=3/2$. They pass through pole-skipping points ({\em open dots}).}
  \label{fig:quasinormalmode}
\end{figure}

We numerically we compute the QNMs of the system as follows. We solve Eqs. \eqref{EomTotalD} for the fields $V_{a,b}$ 
  under infalling boundary conditions at the horizon and sourceless boundary conditions at the asymptotic boundary by setting
$s_{a,b}=0$ in \eqref{eq:asy-vab}. A representative result of this calculation is shown in Fig. \ref{fig:quasinormalmode}. With the Lifshitz scaling exponent fixed at  $z=3/2$, we identify two key lower modes. The gapless hydrodynamic mode (blue) shows the precise subdiffusive  behavior $\omega=-i D_4 k^2$, while  a purely imaginary gapped mode (yellow) follow the dispersion relation $\omega=-i\omega_0-i D k^2$ with $D\simeq D_4$.  Furthermore, both modes pass through  the pole skipping point located at $\big((\omega=-i2\pi T, k/T^{\frac{1}{z}}=2(3- \sqrt{5})^{\frac{1}{2}} \pi^{\frac{2}{3}}/7^{\frac{1}{6}}\simeq 2.71)$ and $(\omega=-i2\pi T, k/T^{\frac{1}{z}}=2(3+ \sqrt{5})^{\frac{1}{2}} \pi^{\frac{2}{3}}/7^{\frac{1}{6}}\simeq 7.10)\big)$. Details of calculation on pole skipping for the transverse fluctuations can be found in App. \ref{app:ps}. The analytical structure of dispersion relation in non-relativistic hydrodynamics differs fundamentally from that in relativistic hydrodynamics  \cite{Grozdanov:2020koi} by comparing the constraints on coefficients in  dispersion relation. 
Our results for the hydrodynamic mode agrees with previous numerical QNM studies in  \cite{Andrade:2022irg}. 
We note that our analytical calculation, focused on the limit $\omega \to 0$, captures the  dispersion relation of the hydrodynamic mode but not that of the gapped mode, whose detailed analytical exploration is left for future work. 

\begin{figure}[!h]
\begin{center}
  \includegraphics[width=0.47\textwidth]{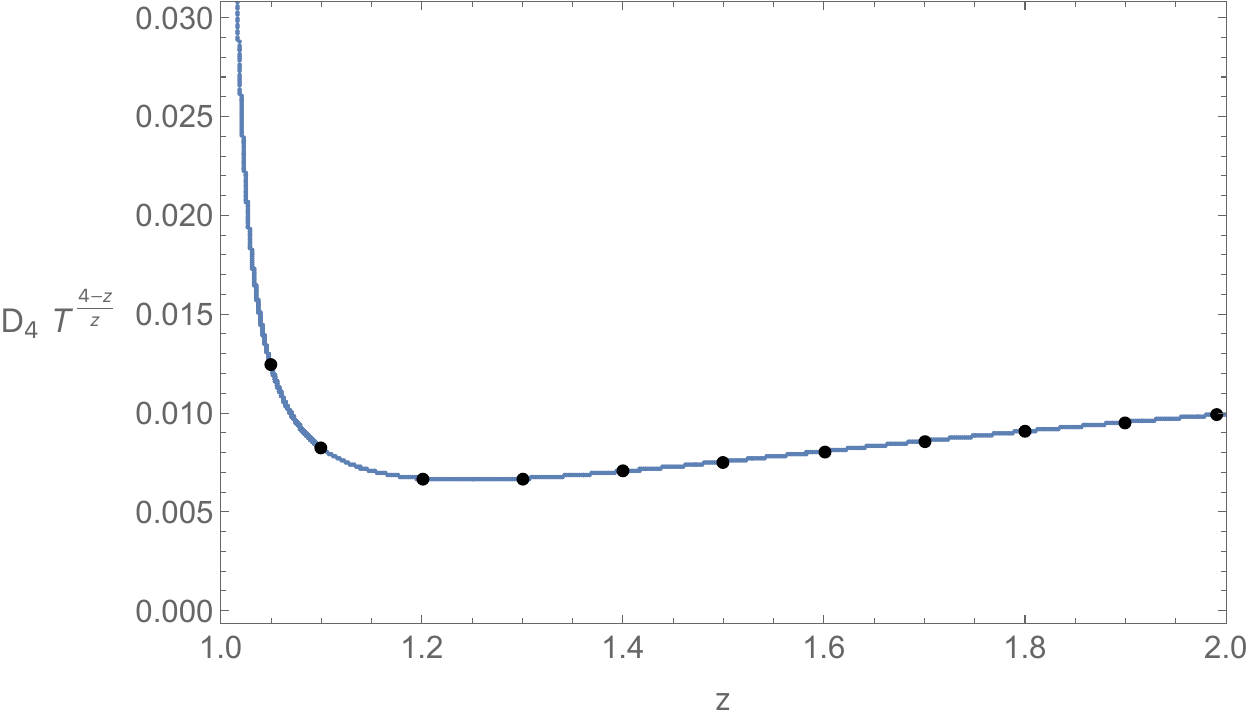}
  \includegraphics[width=0.45\textwidth]{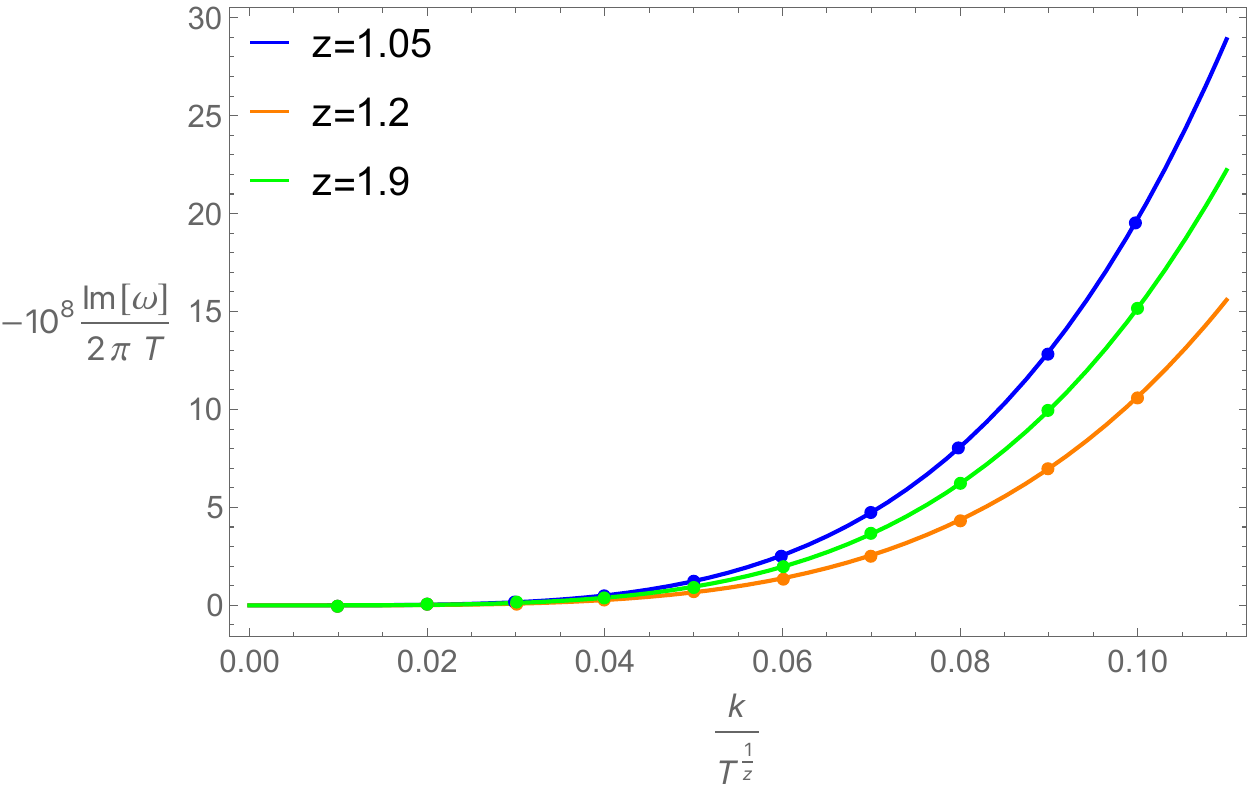}
  \end{center}
  \caption{\small {\em left:
  } The subdiffusive $D_4$ as a function of $z$. The blue dots are from \eqref{eq:D4} while the black dots are from numerical QNM calculation. {\em Right:} The disperssion relations for hydrodynamic modes (dots) at different $z$. The lines are analytic dispersion relations \eqref{eq:dis} with $D_4$ from \eqref{eq:D4}.}
  \label{fig:hydromode}
\end{figure}

In the left panel of Fig. \ref{fig:hydromode}, we plot the  subdiffusive $D_4$ from \eqref{eq:D4} as a function of $z$. 
The right panel shows the QNMs at small $k$ for varies $z$ values, where dots represent numerical results, while the solid line is the analytical curve obtained using the subdiffusive $D_4$ at same $z$  calculated from \eqref{eq:D4}. Our numerical results confirm the analytical dispersion relation \eqref{eq:dis} with the subdiffusive coefficient given by \eqref{eq:D4}. 

\vspace{0.4cm}
\section{Conclusion and discussion}
\label{sec:conclusion}

In this work, we have investigated the low-energy hydrodynamics of strongly coupled non-relativistic many-body systems using Lifshitz holography within the Einstein–Maxwell–dilaton model. The anisotropic Lifshitz scaling symmetry ($  z>1  $) at the ultraviolet fixed point naturally breaks Lorentz boost invariance, yielding a boundary field theory whose geometry is described by torsional Newton–Cartan spacetime. This setup provides a powerful holographic laboratory for exploring non-relativistic hydrodynamics.


Focusing on the shear sector, we analyzed the dynamics of momentum in the hydrodynamic limit. Unlike relativistic hydrodynamics, where energy flux and momentum density are tightly coupled and shear diffusion is universal ($  \omega = -i D k^2  $), non-relativistic systems treat the energy flux $  E_i  $ and momentum density $  P_i  $ as independent observables that both couple to the shear stress tensor $  \pi_{ij}  $. This additional structure fundamentally modifies the low-energy pole structure.


Our key result is the breakdown of standard shear diffusion in favor of subdiffusive momentum transport characterized by the quartic dispersion relation
$  \omega = -i D_4 k^4  $, 
where $  D_4  $ is a subdiffusive constant that depends on an integral over the entire bulk geometry and $z$  (see Eq.~\eqref{eq:dis}). Remarkably, this quartic scaling is universal for all $1<z<2$ and constitutes a genuine signature of non-relativistic hydrodynamics rather than a direct consequence of the Lifshitz scaling itself.
We established this result through a combination of analytic and numerical methods. Using the matching technique, we demonstrated that a leading-order analysis of linear fluctuations is insufficient to capture the hydrodynamic pole; only after iteratively constructing the next-to-leading-order solutions does the generalized matching method correctly reproduce the subdiffusive dispersion. The  numerical computations of the quasi-normal modes independently confirm the presence of these subdiffusive poles, with excellent agreement between the analytic and numerical approaches (see Fig. \ref{fig:hydromode}). Furthermore, we found  that the first non-hydrodynamic mode is purely imaginary and gapped,  following the dispersion relation $\omega=-i\omega_0-i D k^2$, and that both the hydrodynamic and the first non-hydrodynamic modes exhibit  pole-skipping behavior (see Fig. \ref{fig:quasinormalmode}). 
This work provides a new perspective for understanding anomalous transport in strongly correlated  non-relativistic systems.


There are many related interesting questions for the future 
research,  as follows:
\begin{itemize}
\item A natural question is to  understand  the physical origin of the observed subdiffusion from a symmetry or dual field theory perspective. This phenomenon may stem from the fact that, in EMD framework, the dual field theory of the asymptotic Lifshitz spacetime is a thermal fluid in torsional Newton-Cartan geometry, subject to the constraint that the longitudinal gauge field coincides with the spatial shift along the transverse direction. 
\item A complete characterization of the full hydrodynamic theory, including the complete set of constitutive relations and dissipative terms,  would provide a comprehensive description of these systems. Constructing an EFT that couples consistently to TNC geometry would be a natural and powerful next step. 
\item Extending our study to the region $z \geq 2$ would be particularly interesting, as the boundary conditions there are considerably more complex than those considered in our current work.
\item Exploring the potential connection between EMD Lifshitz holography and fracton phases appears particularly intriguing. Recent proposals suggest that finite particle density can spontaneously break Milne boost symmetry, leading to fracton-like behavior \cite{Matus:2024kyg} and even fracton superfluid phases \cite{Jain:2023nbf}. Investigating whether such mechanisms emerge holographically in finite-density Lifshitz systems could bridge holographic methods with the rich physics of higher-rank symmetries and fractonic matter. 
\item It would be interesting to realize superdiffusion with disperssion relation $\omega= -i D_a k^a$ $(a<2)$  from holography. 
\item Generalizing our analysis to hyperscaling-violating Lifshitz black holes \cite{Kiritsis:2016rcb} would allow us to probe how the breakdown of standard Fickian diffusion and the emergence of subdiffusive or superdiffusive momentum transport are affected by additional scaling exponents beyond pure Lifshitz symmetry. 
\item For greater relevance to realistic condensed matter systems, it would be interesting to incorporate momentum dissipation mechanisms \cite{Andrade:2016tbr} and study their impact on the subdiffusive dispersion. 
This could reveal how disorder, pinning, or lattice effects modify the universal quartic scaling we have uncovered.
\item Using a Schwinger-Keldysh path integral approach \cite{Glorioso:2018mmw,Liu:2024tqe, Ahn:2025odk} to derive an EFT for subdiffusive momentum dynamics would offer a systematic way to incorporate fluctuations, noise, and real-time response, potentially unifying our holographic findings with microscopic quantum field theory descriptions.
\item It would be interesting to study the late-time nonlinear effect in Lifshitz holography  \cite{Liu:2025hfv}. 

\end{itemize}
We leave these questions for the future. 

\section*{Acknowledgments}
We would like to thank Fran Pe{\~n}a-Ben{\'\i}tez, Hao-Tian Sun, Ya-Wen Sun for useful discussions. This work is supported by the National Natural Science Foundation of China grant No. 12375041 and 12575046.

\appendix
\section{ Pole skipping of transverse fluctuations in Lifshitz holography}
\label{app:ps}
This appendix details the pole-skipping points for the transverse fluctuations. Pole skipping is a phenomenon first linked to quantum chaos and now recognized as a generic feature of correlators for operators with diverse spins and conformal dimensions
\cite{Grozdanov:2017ajz, Blake:2018leo, Blake:2019otz}. Here we show that pole skipping also exists for operators in non-relativistic holography. 

Here we use the determinant method of Ref.\cite{Blake:2019otz} to obtain the pole-skipping points. After inserting (\ref{eq:nearexp}) into (\ref{EomTotalD}) and expanding the equations in powers of $(r-r_h)$, each equation takes the following near-horizon form:
\be
\begin{split}
S_1=\sum_{n=1}^{\infty} S_1^{(n)}(r-r_h)^{n-1}=0\,, \\
S_2=\sum_{n=1}^{\infty} S_2^{(n)}(r-r_h)^{n-1}=0\,. \\
\end{split}
\ee
The coefficients are computed as 
\be\label{near-horizon equations}
\begin{split}
S_1^{(1)}&=M_{11}\nu_{a}^{(0)} + M_{12}\nu_{b}^{(0)} + 0 + M_{14}\nu_{b}^{(1)}\,, \\
S_2^{(1)}&=M_{21}\nu_{a}^{(0)} + M_{22}\nu_{b}^{(0)} + M_{23}\nu_{a}^{(1)} + M_{24}\nu_{b}^{(1)}\,, \\
S_1^{(2)}&=M_{31}\nu_{a}^{(0)} + + M_{32}\nu_{b}^{(0)} + M_{33}\nu_{a}^{(1)} + M_{34}\nu_{b}^{(1)} + 0 + M_{36}\nu_{b}^{(2)} \,, \\
S_2^{(2)}&=M_{41}\nu_{a}^{(0)} + + M_{42}\nu_{b}^{(0)} + M_{43}\nu_{a}^{(1)} + M_{44}\nu_{b}^{(1)} + M_{45}\nu_{a}^{(2)} + M_{46}\nu_{b}^{(2)}\, , \\
\ldots \\
\end{split}
\ee
where $M_{ij}=M_{ij}(\omega,k^2)$ is determined by the background solutions. The explicit forms of the $M_{ij}$ are rather complicated and we just show the first few terms
\be
\begin{split}
M_{11} &= - \left(\frac{4\pi T}{z+2}\right)^{\frac{3z-1}{z}}(z-4)(z+2)^2 \,,\\
M_{12} &= - \frac{1}{4}\left(\frac{4\pi T}{z+2}\right)^{\frac{z-1}{z}}(z+2) \left( 4k^2 + \left(\frac{4\pi T}{z+2}\right)^{\frac{2}{z}} \frac{\omega}{2\pi T}(z+2)\Big[i(z+3)+\frac{\omega}{2\pi T}(z-1)\Big] \right) \,,\\
M_{12} &= 0 \,,\\
M_{14} &= -i \left(\frac{4\pi T}{z+2}\right)^{\frac{z+2}{z}} (z+2)^2 \frac{\omega}{2\pi T} \left(\frac{\omega}{2\pi T} + i\right) \,,\\
\ldots \\
M_{23} &= - \frac{16i \pi T}{(z+2)^2} \left(\frac{\omega}{2\pi T} + i\right) \left(\frac{\omega}{2\pi T}\right)^{-2} \,,\\
M_{24} &= \frac{4i}{z+2} \left(\frac{\omega}{2\pi T} + i\right) \left(\frac{4\pi T}{z+2}\right)^{\frac{2-z}{z}} \left(\frac{\omega}{2\pi T}\right)^{-2} \,,\\
\ldots \,.\\
\end{split}
\ee
We find that the equations can be written as follows:
\begin{equation}
M(\omega,k^2)\cdot V \equiv
\begin{pmatrix}
M_{11} & M_{12} & 0 & M_{14} & 0 & 0 & \ldots \\
M_{21} & M_{22} & M_{23} & M_{24} & 0 & 0 & \ldots \\
M_{31} & M_{32} & M_{33} & M_{34} & 0 & M_{36} & \ldots \\
M_{41} & M_{42} & M_{43} & M_{44} & M_{45}& M_{46} & \ldots\\ 
\ldots & \ldots & \ldots & \ldots & \ldots & \ldots
\end{pmatrix}
\begin{pmatrix}
\nu_{a}^{(0)} \\ \nu_{b}^{(0)} \\ \nu_{a}^{(1)} \\ \nu_{b}^{(1)} \\ \nu_{a}^{(2)} \\ \nu_{b}^{(2)} \\ \ldots
\end{pmatrix}
=0\,.
\end{equation}
One can easily find that $M_{2n-1,2n+2}=M_{2n,2n+1}=M_{2n,2n+2}=0$ at special frequencies $\omega_n=-i2\pi T n$ with $n=1,2,\cdots $. The locations of the pole-skipping points are given as solutions to the equations
\begin{equation}
\omega_n=-i2\pi T n \,,~~~~ \det(M^{(2n)})=0\,,
\end{equation}
where $M^{2n}$ is an $2n \times 2n$ matrix that corresponds to keeping the first $2n$ rows and $2n$ columns of $M(\omega,k^2)$.

We find for each pole-skipping frequency $\omega_n=-i2\pi T n$, there are $2n$ corresponding pole-skipping wavenumbers ${k^2_n}$, of which only two are positive real values while the rest are complex. Numerical checks up to $n=3$ confirms that for pole-skipping frequency $\omega_n$, the hydrodynamic mode and the first non-hydrodynamic mode respectively  pass through the two real $k_n$ pole-skipping points. This behavior is expected to hold for arbitrary $n$. 

Here we just show that at pole-skipping points of the frequency $\omega_1$, the positive values of ${k^2_1}$ are 
\begin{equation}
{k^2_1}= 2^{\frac{4}{z}-1} \pi^{\frac{2}{z}}(z+2)^{1-\frac{2}{z}}(3 - \sqrt{8z-7})T^{\frac{2}{z}}\,,~ 2^{\frac{4}{z}-1} \pi^{\frac{2}{z}}(z+2)^{1-\frac{2}{z}}(3+\sqrt{8z-7})T^{\frac{2}{z}}\,.
\end{equation}
When $z=3/2$, these points are the locations of circles shown in Fig.  \ref{fig:quasinormalmode}. 


\end{document}